# Model of contact friction based on extreme value statistics

A. Malekan[1] and S. Rouhani[2]

Department of Physics, Sharif University of Technology, Tehran, PO Box 11165-9161, Iran

## Abstract

We propose a model based on extreme value statistics (EVS) and combine it with different models for single-asperity contact, including adhesive and elasto-plastic contacts, to derive a relation between the applied load and the friction force on a rough interface. We find that when the summit distribution is Gumbel and the contact model is Hertzian we have the closest conformity with Amontons` law. The range over which Gumball distribution mimics Amontons` law is wider than the Greenwood-Williamson (GW) Model. However exact conformity with Amontons` law does not seem for any of the well-known EVS distributions. On the other hand plastic deformations in contact area reduce the relative change of pressure slightly with Gumbel distribution. Interestingly when Elasto-plastic contact is assumed for the asperities, together with Gumbel distribution for summits, the best conformity with Amontons` law is achieved. Other extreme value statistics are also studied, and results presented. We combine Gumbel distribution with GW – Mc Cool model which is an improved case of GW model, it takes into account a bandwidth for wavelengths α. Comparison of this model with original GW – Mc Cool model and other simplified versions of the Bush-Gibson-Thomas (BGT) theory reveals that Gumbel distribution has a better conformity with Amontons` law for all value of α. When adhesive contact model is used, the main observation is that for zero or even negative applied load, there is some friction. Asperities with height even less than the separation of two surfaces are in contact. For a small value of adhesion parameter, a better conformity with Amontons` law is observed. Relative pressure increases for stronger adhesion which means that adhesion controlled friction dominated by load controlled friction. We also observe that adhesion increases on a surface with a lower value of roughness.

[1] malekan_azadeh@physics.sharif.edu
[2] rouhani@ipm.ir

# 1-Introduction

Friction between solid bodies is an extremely complicated physical phenomenon, acting on many scales [1-5]. Amontons claimed that Frictional force is proportional to the normal load and is independent of the apparent contact surface, relative velocity and temperature. In other words, there is a linear dependence between normal load and friction force for a wide range of loads and friction coefficient is merely dependent on the material of the two surfaces in contact [6]. Various settings [7-9] were used to test these claims. It seems Amontons` law does not hold completely in all cases. However, to the first order of approximation, the friction law formulated very simply:

$$f = \mu F \qquad , \qquad (1)$$

where $\mu$ is the friction coefficient and $F$ is the normal load. This first order approximation serves many engineering applications. However, its Physical basis remains a mystery. It has been known for a while that many qualifications to this simple relation hold. Coulomb discovered that the static frictional force between two surfaces increases with the contact time [10-11]. Creep process is a possible mechanism which leads to this phenomenon. Because of creep processes, the real contact area grows with time and this growth is faster at higher temperatures [12]. Hence the static frictional force has a logarithmic dependence on time since an increase in the contact area reduces the speed of creep process [13]. The linear dependence of the basic frictional force is not valid for all force domains. Although, the linearity holds for several orders of magnitude of the normal load for metallic materials [14], it breaks down for materials such as polymers and elastomers or soft metals [15]. The frictional force is not completely independent of roughness. It shows a negligible dependence on it. The friction coefficient for extremely smooth metal surfaces is larger than rough surfaces [8]. Also, further deviations from the simple Amontons` law have been observed in rubber which exhibits unusual asymmetry in friction direction [9]. Looking at sliding friction, in the first approximation, the coefficient of friction is independent of speed [6] although experiments show that friction force has some dependency on the sliding velocity. Friction force remains constant for moderate velocities while it decreases for high velocities. For very small velocities, increasing velocity results in an increase of friction force [6]. Various dynamic models were suggested to explain velocity dependence of friction [16-7].

Besides of all of these mentioned deviation from Amontons` law, enormous theoretical efforts have done to substantiate Amontons` claim [17-18]. One of the early explanations of Amontons` law is given by Bowden and Tabor [14]. Actual contact occurs just in the summits because of surface roughness. They considered complete plastic contact and therefore the actual area of contact is connected to hardness indentations. The total area of actual contact $A$ is; $A = F / H$, where $H$ is the hardness of the softer material and $F$ is the normal load. The frictional force is $f = \tau_s F / H$, and local shear stress is $\tau_s$. They propose a coefficient of friction $\mu = \tau_s / H$, as the ratio of two material properties.
Since the real surfaces are rough on the microscopic scale, contact occurs in summits of asperities. GW model proposed an elastic and adhesion-less asperity contact with Gaussian distribution for summit`s heights. They found an approximately constant pressure during loading [19].Archard simulated a rough surface as a series of spheres which are superimposed hierarchically [20]. He proved that the relation between the real contact area $A$ and the normal load $F$ is given by a power law $A \sim F^\alpha$ where in the case of a complex real surface, the exponent α ≈1. A is nearly proportional to the load, according to Amontons` law.  Bush-Gibson and Thomas [21](BGT) used a statistical theory of isotropic randomly rough surfaces, which utilizes a bandwidth parameter. They used Longuet-Higgins [22] and Nayak [23] probability distribution of summits for surface statistics of isotropic surface:

$$P(z, R_1, R_2) = \frac{\sqrt{27}}{(4\pi)^2} \frac{1}{m_2 m_4 \sqrt{m_0 m_4}} C_1^{1/2} exp\left[-C_1\left(\frac{z}{m_0^{1/2}} + \frac{3}{2\sqrt{\alpha}}\frac{1/R_1+1/R_2}{m_4^{1/2}}\right)^2\right] \left|\left(\frac{1}{R_1}\right) - \left(\frac{1}{R_2}\right)\right| \times (R_1 R_2)^{-1} exp\{-\frac{3}{16 m_4}[3(1/R_1 + 1/R_2)^2 - 8(R_1 R_2)^{-1}]\} \qquad (2)$$

Power spectral density (p.s.d.) is the Fourier transform of height autocorrelation function for a Gaussian and isotropic surface, $z$ is the summit height and $R_1$, $R_2$ are summit radius. The zero, second and fourth moments $m_0$, $m_2$, $m_4$ of the surface roughness power spectrum, are functions of the breadth of the surface roughness and wavelength $\alpha = \frac{m_0 m_4}{m_2^2}$.Longuef and Higgins have shown in a random and isotropic surface $\alpha \geq 3/2$. The p.s.d. spreads by increasing α. In BGT theory an isotropic rough surface with joint summit and curvature distribution is assumed by Longuet-Higgins and Nayak [23]. This surface is taken in contact with a flat surface. The spheres of the GW model are replaced by paraboloids.  The contact area A turns out to be proportional to the normal load [21] provided that normal applied load is very low or A is well below the apparent area of contact.
Persson [24] linked the apparent contact area $A$, to a length scaleΛ. Here the length Λ is the projection of the contact area when the original surface considered is smooth on all length scales below Λ. The ratio $\xi = L/\Lambda$ is the magnification of the surface, where L is the length of the sample. Persson assumed $P(\sigma, \xi)$, the stress distribution at the magnification $\xi$, satisfies a diffusion-like equation. He also found a linear relationship between normal load and the real area of contact provided that the normal applied load is small.
In this paper, we propose a model for friction based on extreme value statistics (EVS) [25]. The rough contact friction force is given by two considerations, what the model for asperity contact is and what is the summit`s distribution. The simplest choice for single-asperity contact is an elastic contact model or the Hertzian asperity [26]. The others are adhesive and elastic-plastic contacts models, the Maugis- Dugdale (MD) [27] model is a general adhesive theory and Johnson–Kendall–Roberts (JKR) and Derjagin-Muller-Toropov (DMT) are its limiting cases. Another option is the Chang- Etsion- Bogy (CEB) [28] or elastic-plastic model based on volume conservation of an asperity during plastic deformation. We use EVS for independent and identically distributed (IID) variables and the maximum height $h_m$ (1+1) Kardar-Parisi-Zhang (KPZ) model. We follow the GW model assumptions (see below) and combine the various possibilities of asperity contact and EVS distributions and solve numerically to obtain a relationship between contact area, friction force and the applied load for a variety of distributions and contacts. Since there is no direct evidence for what is to be used for the EVS distribution we test the various universal EVS distributions to see which produces a better Amontons law. In addition we use EVS for summit`s distribution in some simplified version of BGT models which consider a wavelength for summit`s radius. The conclusion being Gumbels' distribution, with an elasto-plastic contact. It has to be emphasized here that surface correlations are ignored in this kind of analysis, with the exception of the KPZ surface.

The roadmap of this paper is as follows. In section 2 we describe the Greenwood-Williamson (GW) Model which sets the basis of our analysis. In section 3 we provide a very brief introduction to extreme value statistics (EVS). In Section 4 we combine EVS with single asperity models and numerically calculate the contact pressure for a number of universal EVS's and different asperity models. In section 5 we try to question the assumptions of the GW model. We close by some concluding remarks.

| Nomenclature | |
|---|---|
| $f$ = friction force | $E^*$ = effective Young`s moduli |
| $F$ = normal load | $N_0$ = total number of asperities |
| $\mu$ = friction coefficient | $\alpha$ = bandwith of wavelengths |
| $\bar{P}$ = dimensionless force in MD model | $R_G$ = Greenwood model`s radius |
| $p$ = pressure | $R_A$ = NT`s model radius |
| $A$ = real contact area | $R_{1,2}$ = asperity raduis |
| $A_0$ = nominal area of contact | $\omega$ = surface roughness |
| $\bar{A}$ = dimensionless area in MD model | $\alpha_k$ = the zeros of Airy function |
| $H$ = Hardness of the softer material | $U(a,b,c)$ = confluent hyper geometric function |
| $\tau_s$ = shear stress | $d_{max}$ = the highest summit limit |
| $\sigma$ = contact stress | $\eta$ = density of asperities |
| $\sigma_{adh}$ = adhesion stress | |
| $m_0$ = zero moment | |
| $m_2$ = second moment | $t$ = dimensionless distannce $\dfrac{d}{\sqrt{m_0}}$ |
| $m_4$ = fourth moment | |
| $\Lambda$ = length scale | $\delta$ = interference of two surfaces |
| $L$ = length of sample | $\delta_c$ = critical interference |
| $\xi$ = magnification of the surface | $\psi$ = plasticity index $\Delta\gamma$ = surface energy |
| = magnification of the surface | $\Delta\gamma$ = surface energy |
| $P(\sigma,\xi)$ = stress distribution at $\xi$ | $z_0$ = the equilibrium in Lennard – Jones force |
| $E_{1,2}$ = Young`s moduli | $\lambda$ = adhesion parameter |
| $\nu_{1,2}$ = Poisson's ratios | |
| $d$ = sepration between two surfaces | |

## 2- Greenwood-Williamson Model

Greenwood and Williamson [19] developed a theory based on Hertz contact theory, assuming a flat rigid plane in contact with a rough surface where the distance between the flat rigid planes from the mean height of rough surface is d. All asperities have the same radius $R$. The height of the peaks is stochastically distributed around an average value figure 1. If $\Phi(z)$ is the summit distribution and there are $N_0 = \eta\, A_0$ asperities within a nominal area of $A_0$, the total real area of contact is:

$$A = \pi N_0 \int_d^\infty dz\, \Phi(z)\, R\, (z - d) \qquad (3)$$

And total load is the summation of loads of every single asperity in contact:

$$F = \frac{4}{3} N_0 E^* \int_d^\infty dz\, \Phi(z)\sqrt{R}(z - d)^{3/2} \qquad (4)$$

Where $E^*$ and R defines as:

$$\frac{1}{E^*} = \frac{1-v_1^2}{E_1} + \frac{1-v_2^2}{E_2}, \tag{5}$$

$$\frac{1}{R} = \frac{1}{R_1} + \frac{1}{R_2}, \tag{6}$$

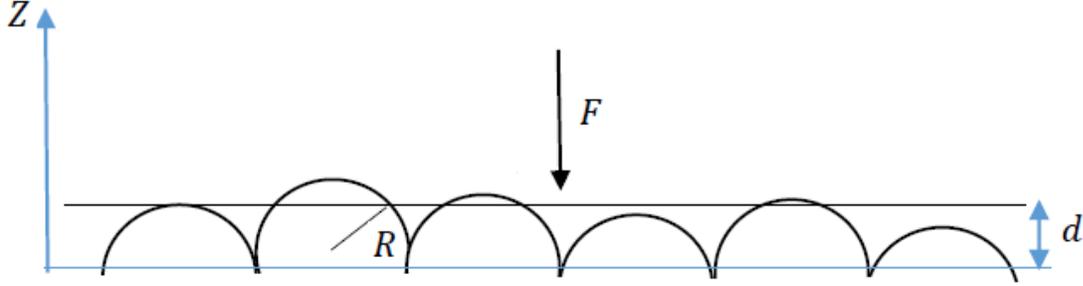

Figure 1.The schematic drawing of GW model.

Assume a Gaussian distribution for summits of asperities [19].

$$\Phi^*(z) = (\frac{1}{2\pi\omega^2})^{\frac{1}{2}} \exp\left(-\frac{z^2}{2\,\omega^2}\right) \tag{7}$$

It is better to use the natural length scale of the problem namely the roughness $\omega$ (RMS of the width of the height of asperities) as a dimensional quantity. And the real area of contact becomes:

$$A = \pi\,\omega\,N_0 R \int_{\frac{d}{\omega}}^{\infty} dz\ \Phi(z) \left(\frac{z}{\omega} - \frac{d}{\omega}\right) \tag{8}$$

The total load after scaling with roughness obtained:

$$F = \frac{4}{3} N_0\,E^* \omega^{3/2}\ R^{1/2} \int_{\frac{d}{\omega}}^{\infty} dz\ \Phi(z)\,(\frac{z}{\omega} - \frac{d}{\omega})^{3/2} \tag{9}$$

And $\Phi(z) = \omega \Phi^*(\omega\,z)$. They plot the load divided by the actual area of contact $\frac{F}{A}$ versus the surface separation when $\varphi(z)$ is Gaussian distribution (figure 2). If we assume that the actual area of contact is proportional to the friction force, this plot should give us the friction coefficient.

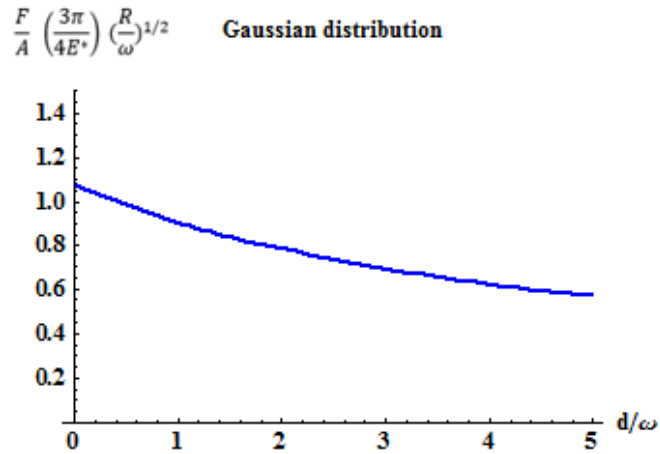

*Figure2. The pressure of contact as separation of surfaces when the Gaussian distribution is summits` distribution. Clearly, the friction coefficient is not independent of the load, though there is the almost constant behavior for the range of (2.5ω -5ω).*

## 3-Extreme value statistics

The assumption that the summit distribution is Gaussian is too simple in GW model. Given a height distribution, we need to look at summits distribution since the asperities are in contact in their summits. This distribution described by extreme value statistics (EVS).

EVS is a branch of statistics which strives to find the probability distribution of maxima and minima of given distributions. Given a random height distribution, we are seeking distribution of its maxima. This is given by the extreme value statistics of ϕ(x). EVS has a lot of applications in many natural phenomena and engineering [29-32], and it might be a proper choice for summits distribution. Unfortunately, EVS of any given height distribution is not known, but it is known for some special cases. Here we will zoom into these special cases and make some guesses for the more general cases. Based on the mother distribution, there are three types of universal limit distributions for independent and identically distributed (IID) and for a large number of random variables. This is known as the Gnedenko's classical law of extremes [33]. The PDF of maxima is given by Fisher-Tippett-Gumbel distribution [34] when the distribution of IID variables has tails decaying faster than power law but unbounded such as $P(x) \sim e^{-x^\delta}$ with δ>0.

$$f_1(z) = e^{-\frac{x}{\omega} - e^{-\frac{x}{\omega}}}, \quad \omega > 0, x \in (-\infty, \infty) \tag{10}$$

The Gumbel universality class corresponds to exponential or a Gaussian or gamma distribution for variables. It describes extreme wind speeds, sea wave heights, floods, rainfall, etc. Also, it has applications in size phenomena such as the size of material flaws and surface imperfections and to event magnitudes like queue length, order lead time [29].

For IID random variables, with parent distribution of power law convergence $P(x) \sim x^{-(1+\beta)}$ with $\beta > 0$. The PDF is Fréchet distribution given by:

$$f_2(x) = \frac{\beta}{\omega}(\frac{x}{\omega})^{-\beta-1} e^{-\frac{x}{\omega}^{-\beta}}, \qquad \omega > 0, \; x \in [0, \infty) \tag{11}$$

The Fréchet domain has distributions with an infinite yet heavier tail than the exponential distributions. It corresponds to EVS of Cauchy or Preto distributions. The Fréchet distribution has application in extreme events such as annually maximum one-day rainfalls and river discharges [33]. The maximum loads which engineering devices can tolerate are needed in their service mission [30]. Natural phenomena such as floods, snow accumulation, wave forces, earthquakes, wind pressure, and so forth often caused these loads [29]. Fréchet distribution intrinsic longer upper tail leads to an upward data fit. Therefore, the Fréchet distribution is another alternative for modeling maximum extreme value phenomena in addition to the Gumbel distribution.

For the parent distributions with the bounded tails such as $P(x) \xrightarrow{x \to a} (a-x)^{\beta-1}$ with $\beta > 0$. The PDF is the Weibull distribution:

$$f_3(x) = \beta(\frac{x}{\omega})^{(\beta-1)} e^{-(\frac{x}{\omega})^\beta}, \qquad \omega > 0, x \in [0, \infty) \tag{12}$$

Distributions in this universality class have lighter tails than exponential, which owns a finite upper bound. There are several thousand papers about the Weibull distribution applications in some natural phenomena such as wind-speed data analysis [31], earthquake magnitude [32] and volcanic occurrence data and so on.

On the other hand, there are many distributions which do not belong to the three mentioned domains of attraction. For example, EVS of geometric and Poisson distributions cannot be determined by the standard extreme value distributions. Although, EVS domains of attraction includes most applied distributions, such as Pareto-like distributions (Cauchy), normal and Beta distributions [35].

A general theory similar to those for IID does not exist for strongly correlated random variables. There are few examples, such as maximum heights of a fluctuating (1+1) dimensional interface, where the EVS of a strongly correlated system was computed exactly. Majumdar [36-37] found the PDF of maximum height $h_m$ (1+1) KPZ model has the scaling form for all Lω.

$$P(h_m, L\omega) = \frac{1}{\sqrt{L\omega}} f(\frac{h_m}{\sqrt{L\omega}}) \tag{13}$$

The scaling function named as Airy distribution:

$$f(x) = \frac{2\sqrt{6}}{x^{\frac{10}{3}}} \sum_{k=1}^{\infty} e^{\frac{-b_k}{x^2}} b_k^{\frac{2}{3}} U(\frac{-5}{6}, \frac{4}{3}, \frac{b_k}{x^2}) \tag{14}$$

Where $\omega$ is the surface roughness, $L$ is the length of sample, $b_k = \frac{2}{27}\alpha_k^3$ where $\alpha_k$`s are the absolute value of the zeros of Airy function and $U(a,b,z)$ is a confluent hyper geometric function of the second kind.

## 4-Extreme value Statistics Model of Friction

Let us now repeat the GW model with EVS distributions as $\Phi(z)$. Furthermore we will take various asperity contacts and combine them with EVS. Not knowing what the distribution of the heights is we do not know what the relevant EVS distribution is. Therefore we shall report results for the three universal EVS distributions here.

In fact, at separation $d > d_{max}$, the two surfaces are no longer in contact hence the normal load vanishes. Since all EVS distributions except Fréchet fall quickly the integral of load and contact area converge and $\frac{d_{max}}{\omega}$ replaced by infinity. We plot dimensionless pressure $\frac{4}{3}(E^*/\pi) * (\omega/R)^{1/2} * (F(d,R)/A(d,R))$ as a function of d/ω. Decreasing the normal load and real area of contact with increasing d give us hope that the ratio may turn out approximately constant. Consequently, the Amontons` law concluded, alas this is not the case. The linear relationship between the real area of contact and applied load expected in this interval. Fréchet distributions with $0 \leq \beta \leq 2$ have a fat tail, thus to find the total load and real area of contact we need to set an upper limit to the peak height $d_{max}$. It is natural to assume that the bigger the area of the sample it is likely to encounter a higher maximum peak. How this maximum scales with sample size relates to how the distribution falls at large values. Here we will assume that it scales with the nominal area, $d_{max} \sim A_0^\gamma$. Fréchet distribution scales with size as $N_0^{\frac{1}{\beta}}$, thus $N_0$ is proportional to $A_0$. We need to cut off at a maximum height, so let`s choose $d_{max}$ such that 99 % of summits included hence $d_{max} \sim A_0^{\frac{1}{\beta}}$. We plot the nominal friction force as a function of separation for the Fréchet distribution. The resulting friction force doesn`t depend on $d_{max}$ or equivalently on the nominal area of contact. The friction does not show a monotonous trend in figure 3a. In Fréchet distribution with $\beta > 2$ by increasing load, the number of short summits is not enough to reduce or balance the pressure in high loads figure 3b. Figure 4 shows the pressure of contact when other EVS used as summit distribution. Gumbel distribution has the most uniform pressure in the physical contact condition and shows the best conformity with Amontons` law, Figure4a. Gumbel domain of attraction belongs to mother distributions with exponential decay such as Gaussian. This result is consistent with observations which suggest Gaussian distribution as asperities` heights distribution.

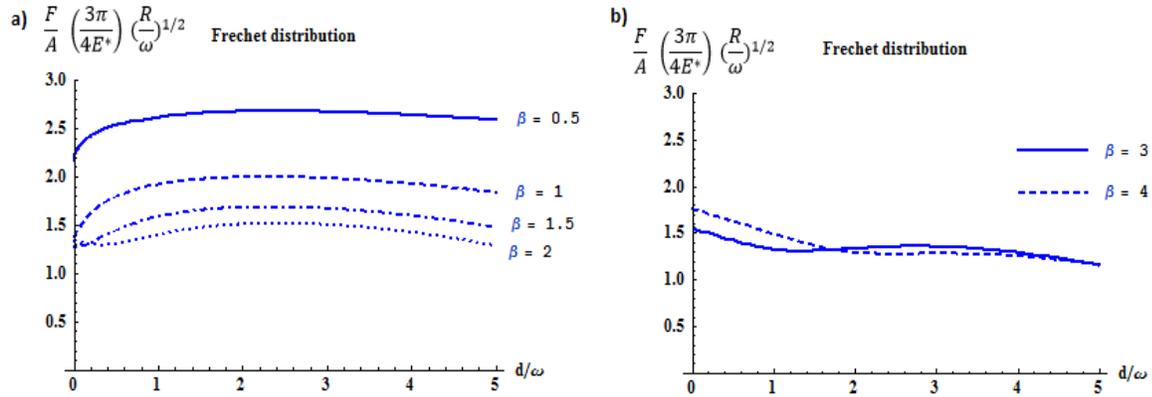

*Figure3. The pressure of contact for the Fréchet distribution and Hertzian contact. a) The pressure of contact does not have a monotonous trend by increasing load when for β<2. b) for β≥2 The pressure of contact has a rise in pressure due to a reduction in the number of short summits for larger separations.*

For $\beta = 1$ Weibull distribution is the Exponential distribution which independent of the particular surface model shows exact proportionality between the load and the area of contact figure 4b, although it is not a fair approximation of the asperities of the surfaces since this means uniform distribution for asperities height [19].

Simplified EVS distribution for 1+1 KPZ model (Airy distribution not to be confused with airy function) for a surface gives [36-37]:

$$\varphi(z) = \begin{cases} 0.15 * |a_1|^2 \, z^{\frac{-10}{3}} e^{\frac{-2a_1^3}{27z^2}} & z \leq 0.56 \\ 1.84 \, e^{-6z^2} & z > 0.56 \end{cases} \quad , \tag{15}$$

where $a_1$ is the first zero of the Airy function. In figure 5 Airy distribution is summit`s distribution with Hertzian contact. It seems Airy distribution is not a good candidate for summit`s distribution since the pressure changes are bigger than other EVS and even Gaussian distribution. In table 1 we observe the range of total load and real area of contact when Gumbel distribution is used as summits` distribution for two surfaces by $E_1 = E_2 = 17 \, GPa$ and $v_1 = v_2 = 0.15$. Observe that relative pressure change in the interval of (ω,5ω) is 0.02 for Gumbel distribution. These values are according to the experimental observation of Nuri and Hailing (1975) [38]. Here we observe that the typical loads are reasonable when our model lies in good conformity with Amontons` law. We therefore observe that Gumbel distribution is best suited and fits the Amonton law.

However elastic contact conserves energy and cannot be a good candidate for friction, thus a plastic component to the asperity behaviour is necessary.

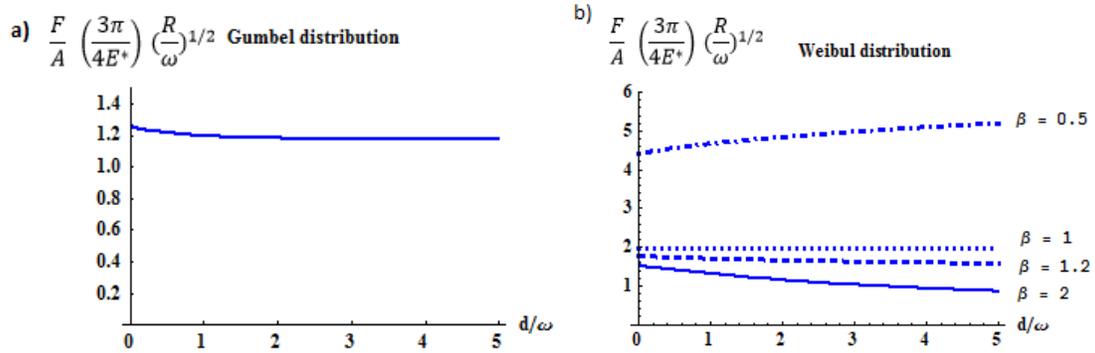

Figure4. Different Weibull and Gumbel distributions as summits distribution with Hertzian contact. a) Gumbel Pdf gives an almost flat pressure which indicates the validity of Amontons` law for the range of (ω,5ω). b) The Weibull distributions have different trends for various values of beta. β=1 corresponds to the exponential distribution which shows exact proportionality between the load and the area of contact independent of the surface model. There is also good correspondence with Amontons Law values of beta near one(6b).

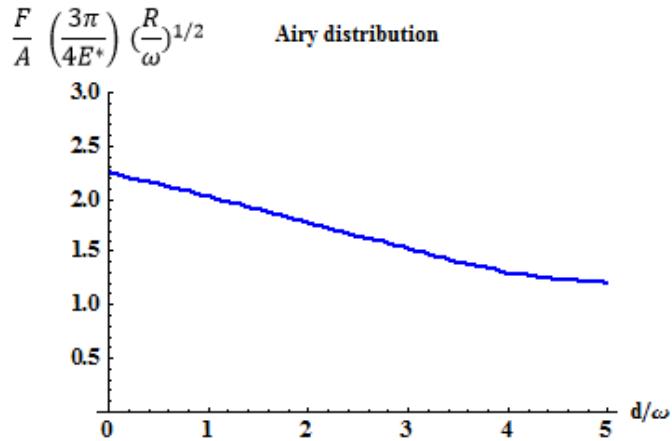

Figure5. Airy distribution has used as summits` distribution with Hertzian contact. In comparison with other EVS distributions and even Gaussian distribution, it has the biggest variation in pressure.

| $\eta R\omega$ | $\dfrac{\omega}{R}$ | $F(5\omega) - F(\omega)$ <br> Kg | $\dfrac{A_{re}(d_2)}{A_0} - \dfrac{A_{re}(d_1)}{A_0}$ |
|---|---|---|---|
| 0.0302 | $8.75 * 10^{-5}$ | 0.0005 − 80 | 0.02 − 0.0001 |
| 0.0374 | $2.00 * 10^{-4}$ | 0.9 − 150 | 0.02 − 0.0001 |
| 0.0601 | $1.77 * 10^{-3}$ | 4.5 − 710 | 0.03 − 0.0002 |
| 0.0401 | $2.48 * 10^{-4}$ | 1.0 − 177 | 0.02 − 0.0001 |

*Table 1. The total load and real area of contact for concrete with Gumbel summits' distribution in $\omega$ and $5\omega$*

Here, we use CEB model [28] of elastic-plastic contact which based on volume conservation of plastically deformed region of the asperity. Figure 6a compares pressure for elastic – plastic model with the Hertzian model. As the plastic index increases, the pressure reduces. Increasing real area of contact due to Plastic deformation, this makes the pressure more uniform. Figure 6b shows the relative change of pressure for fully elastic and elastic- plastic cases with different plastic index. We observe that plastic contact and Gumbel distribution produce the closest result to Amontons` law.

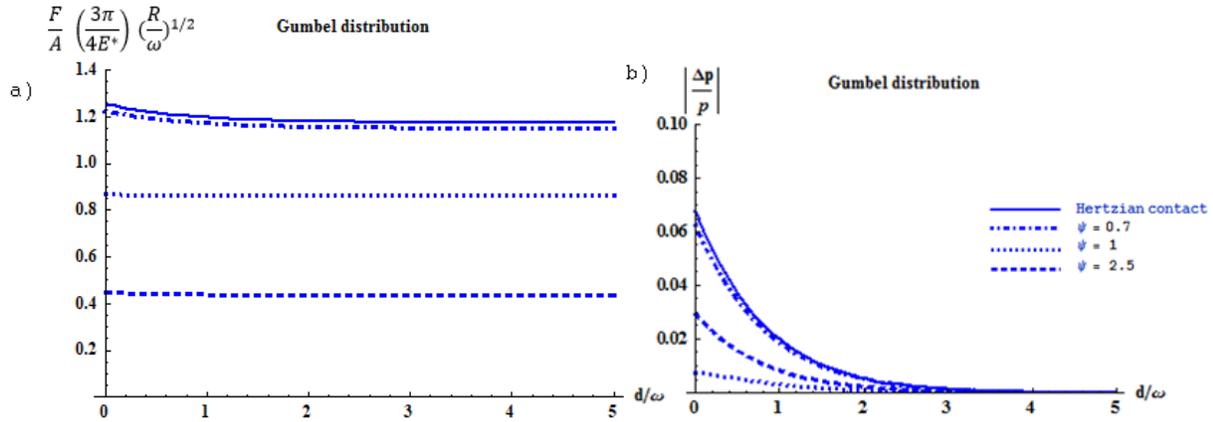

*Figure 6. a)Pressure via separation and Elastic- plastic model as contact asperity. Variation of plasticity index changes the pressure since the real area of contact increases. b) Relative change of pressure via separation. A fully elastic model such as Hertz model has the most pressure changes compared to elastic- plastic contacts*

Maugis [27] introduced two dimensionless parameters $\bar{P} = \dfrac{F}{\pi R \Delta \gamma}$ and $\bar{A} = \dfrac{a}{\left(\dfrac{3\pi \Delta \gamma R^2}{4E^*}\right)^{1/3}}$ for force and area and an adhesion parameter $= 2\, \sigma_{adh} \left(\dfrac{9R}{16\pi \Delta \gamma E^{*2}}\right)^{1/3}$, where $\sigma_{adh}$ is adhesion stress defined as:

$$\sigma_{adh} = \dfrac{\Delta \gamma}{h_0} \quad . \tag{16}$$

Here $\Delta \gamma$ is the surface energy and $z_0$ is the equilibrium in Lennard – Jones force, typically around $1\,Å$ [8].

$$h_0 = 0.97 z_0 \tag{17}$$

If $\lambda > 5$ the JKR analysis becomes appropriate and when $\lambda < 0.1$, the DMT model is applicable. In the intermediate range $0.1 < \lambda < 5$ the MD model has to be applied. For adhesive contact problem, in DMT limit the pressure value is very close to the pressure in the Hertzian model. By increasing the adhesion parameter transition from DMT to JKR limit occurs and the pressure value reduces consequently figure7.

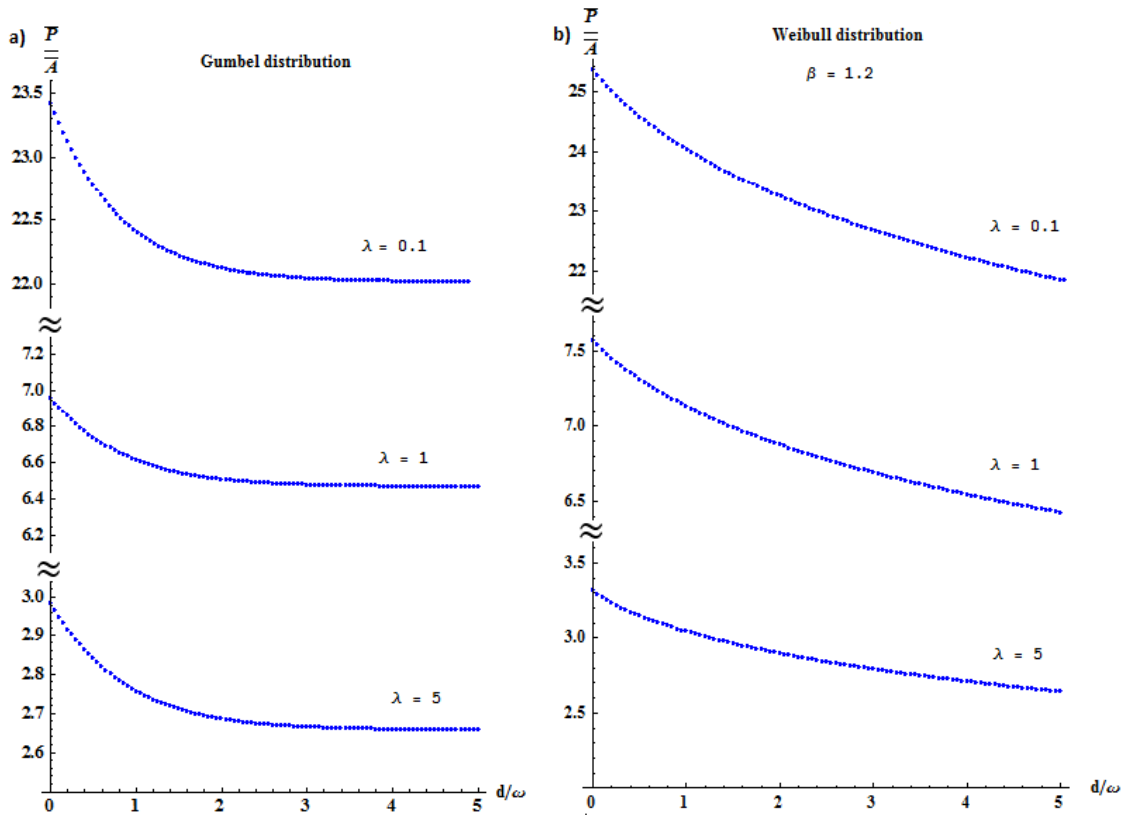

*Figure 7.a) Gumbel distribution: pressure versus separation for contacts with different values of the adhesion parameter λ. For adhesive contacts, pressure has a lower value. b) Weibull distribution β=1.2, pressure versus separation for contacts with different values of the adhesion parameter λ.*

Considering adhesion with MD model in contact, results show asperities of heights even $z < d$ can be in contact. Asperities with height $z > d$ compressed and those with height $d - \delta_c < z < d$ stretched. $\delta_c$ is the separation in which two surfaces takes apart with stretching out two surfaces. The effect of this pull off force is considerable in high value of $\lambda$ figure 8.

Deviations from Amontons` law have been observed by increasing adhesion in materials [39]. Table 2 shows the results of Gumbel distribution for three values of the adhesion parameter for a fixed value of roughness, where we observe an increase in pressure with adhesion. We see more deviation from Amontons` law for higher adhesion parameter. For a 10 times smoother surface with λ=5, the relative pressure change is 0.22 which is 10 times bigger than a rougher surface.

The Amontons-like behavior is dominant with low adhesion parameter. Amontons` law doesn`t describe friction behavior in zero or negative applied load. Adhesive control friction happens in a higher value of adhesion parameter or a smoother surface. When roughness reduces 10 times in a fixed adhesion parameter, the pressure of contact falls significantly figure 9.

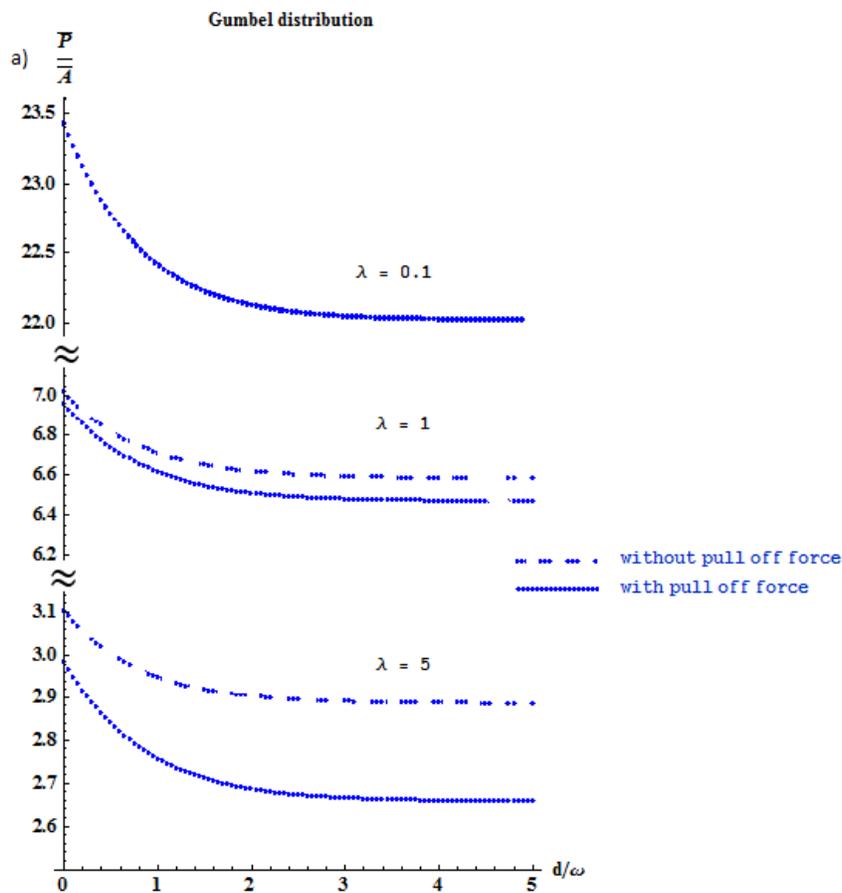

*Figure8. In adhesive contacts, asperities with heights less than the separation of two surfaces are in contact. The pull off force is negligible for small adhesion parameters and it is more effective in high adhesion parameter.*

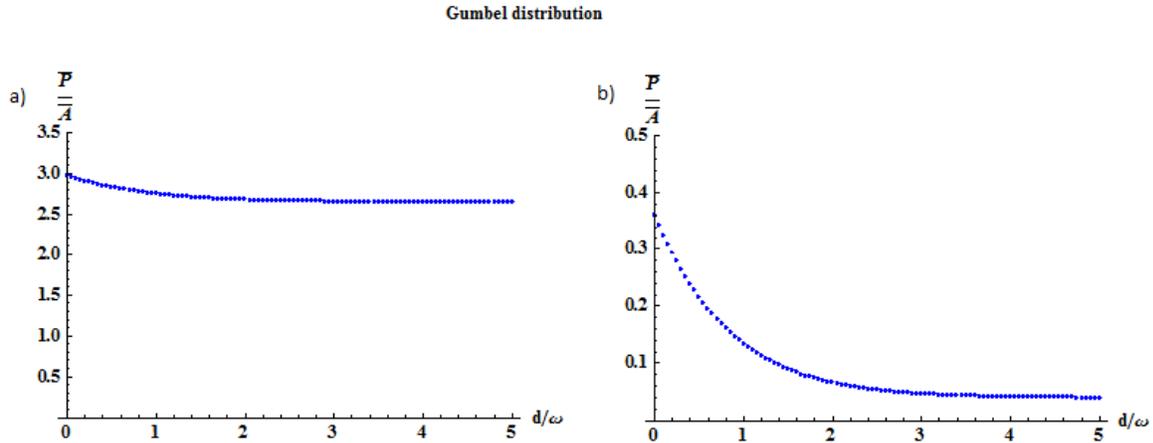

Figure 9. Surface roughness decreases the adhesion effects in contact. a) The pressure of contact for λ=5 in a fixed roughness. b) The roughness of the surface decreased by 10 with λ=5 and result in considerable reduction in the pressure.

| λ | $\frac{p(\omega) - p(5\omega)}{p(5\omega)}$ |
|---|---|
| 0.1 | 0.013 |
| 1 | 0.015 |
| 5 | 0.022 |

Table 2. The pressure change in contact for Gumble distribution with different adhesion parameters. $p(d) = \frac{P(d)}{A(d)}$ is the dimesionless pressure in separation d, and ω is surface roughness.

## 5-Beyond the GW model

Whilst in GW model an identical radius for all asperities is assumed, some authors have tried to extent this to more realistic setting by combining (simplifying) the BGT model with the GW model. In BGT model a parameter $\alpha$ is defined which appears in Longuet-Higgins [22] and Nayak [23] probability distribution of summits for surface statistics of isotropic surface, it is defined as $\alpha = \frac{m_0 m_4}{m_2^2}$ where $m_0$, $m_2$, $m_4$ are the zero, second and fourth moments of the surface roughness power spectrum. The parameter $\alpha$ is an indication of how broad is the distribution of radiuses of asperities.

For instance, Greenwood presented a simplified version of BGT model in 2006 [40]. In this model, the summits are spheres with a distribution for the mean radius $R_G = \sqrt{R_1 R_2}$. Another model presented in [41] the mean radius of a summit is taken as $R_A = \frac{2R_1 R_2}{R_1+ R_2}$, we called it the NT model. An improved model of GW model is Mc Cool [42] who combines GW model and some results of NT statistical model. We refer the reader to [43] for a detailed description of these models.

In order to test our proposal, we use Gumbel distribution as summits distribution in GW-Mc Cool model. Compression of the results with other asperity models presented in [43] shows a better conformity with Amontons` law. Also, the real area of contact has more realistic values compared with models in [43] (figure10). In [43] heights and separation are scaled by surface height variance $\sqrt{m_0}$ instead of summit `s height variance. The relationship between $m_0$ and $\omega^2$ were found by Bush et al [44] as:

$$\omega^2 = (1 - \frac{0.8968}{\alpha})m_0 \qquad (18)$$

The summit variance approaches the surface variance as $\alpha$ becomes large. Table 3 shows the real area of contact when separation is $(\sqrt{m_0} - 5\sqrt{m_0})$ and table 4 shows the pressure change in this distance. For a higher value of $\alpha$ Gumbel distribution has the most uniform pressure which means the closest similarity with Amontons` law in quite a realistic area of contact.

| A | Greenwood 2006 | NT | GW- Mc Cool | EVS |
|---|---|---|---|---|
| 2 | $10^{-7}$–0.061 | $2\times10^{-8}$–0.010 | $3\times10^{-8}$–0.062 | 0.0001–0.085 |
| 10 | $10^{-7}$–0.059 | $1\times10^{-7}$–0.051 | $1\times10^{-7}$–0.051 | 0.0004–0.087 |
| 100 | $10^{-7}$–0.101 | $9\times10^{-8}$ –0.087 | $9\times10^{-8}$-0.079 | 0.0010–0.167 |

Table 3. The real area of contact to the nominal area when the distance between two surfaces is $(\sqrt{m_0} - 5\sqrt{m_0})$.

| | α = 2 | α = 10 | α = 100 |
|---|---|---|---|
| Greenwood 2006 | 0.129 | 0.373 | 0.537 |
| NT | 0.198 | 0.419 | 0.544 |
| GW- Mc Cool | 1.124 | 0.722 | 0.630 |
| EVS | 0.164 | 0.047 | 0.026 |

Table4.The pressure change in the distance $(\sqrt{m_0} - 5\sqrt{m_0})$ for different models. More realistic surfaces have a high value of α.we have the closest result to Amontons` law with Gumbel distribution in GW- Mc cool distribution.

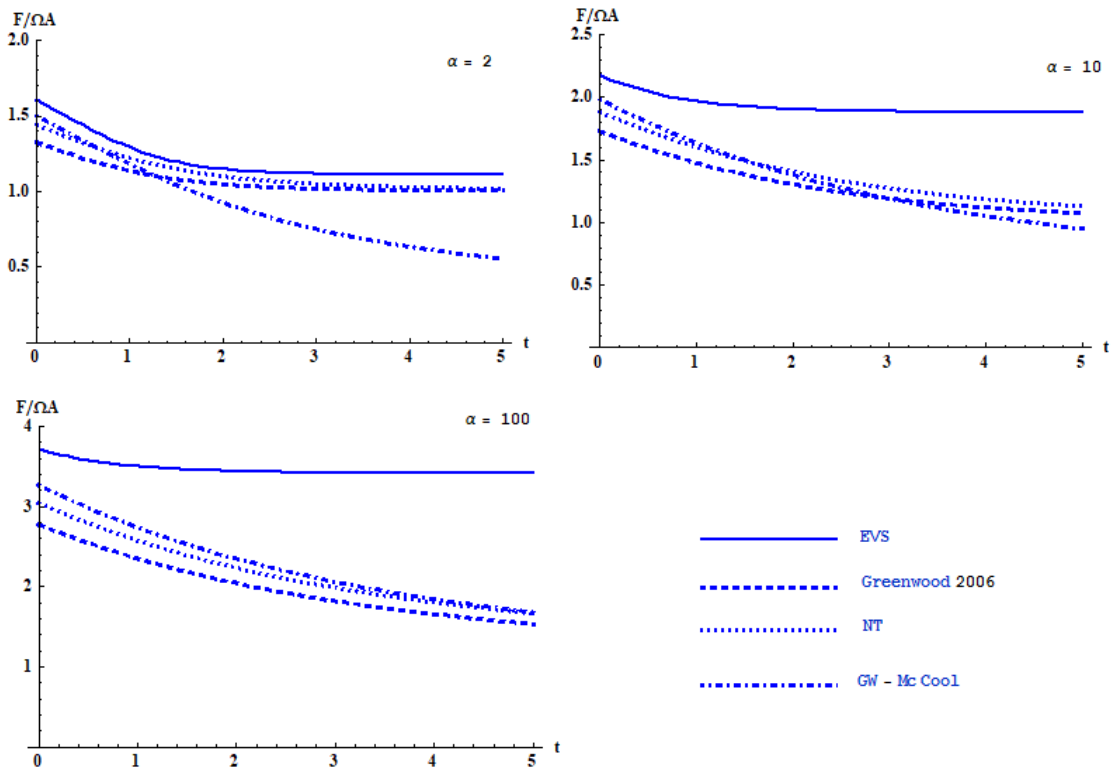

*Figure10. When we use Gumbel distribution in GW- McCool model the value of pressure is more uniform than other models.*

## 6-Conclusions

Amontons` law states a linear relationship between applied load and friction force. It is quite acceptable to take a linear relationship between the friction force and the real area of contact. Therefore the pressure of contact should remain constant.

In this work, we combine various models of single asperity contact, including Hertzian, elastic- plastic and adhesive contact models with Extreme Value Statistics (EVS) for summits` distributions to verify Amontons` law.

Within EVS theory there exist three kinds of universal distributions for independent and identical distributed (IID) variables, namely Fréchet, Gumbel and Weibull distributions. Asperities` height in a real surface is strongly correlated. Here, we consider them as IID variables and use Fréchet, Gumbel and Weibull distributions as summits` height. Surface roughness is a strongly correlated systems and very little is known about EVS of correlated heights. For (1+1)-dimensional KPZ surface, exact EVS distribution is Airy distribution. We extend Airy distribution to the two-dimensional surface and use it as summits` distribution. The resulting pressure varies considerably by the applied load.

Among EVS distributions, Gumbel distribution shows the best conformity with Amontons` law for Hertzian contact. To determine the relevance of the Gumbel distribution one way is to measure the height profile and determine the statistics of the height distribution.

The pressure is almost constant in a relatively large interval of the applied load. Although Weibull distribution with $\beta = 1$ is the Exponential distribution and has a constant pressure with all applied load, it is not a proper candidate for summit distribution since height distribution is uniform for $\beta = 1$. Fréchet distributions with $\beta \leq 2$ are fat tailed. They decay very slowly. For $\beta > 2$ the number of short summits is not enough to reduce or balance the pressure in high loads and therefore the pressure increases. We also combine Gumbel distribution with GW – Mc Cool model which is an improved case of GW model. Here a bandwidth for wavelengths α is assumed. Comparison of this model with the original GW –Mc Cool model and other simplified versions of BGT reveal that Gumbel distribution has a better conformity with Amontons` law for all values of $\alpha$.

The other point of contention is what is the best model for an asperity. Plastic deformations occur during contact. The pressure`s changes is minimum with a combination of plastic and elastic deformations. When adhesion exists in contact, the main observation is friction force in the zero or even negative applied load. Asperities with heights even less than the separation of two surfaces are in contact. For a small value of adhesion parameter Amontons-like behavior is dominant. The adhesion controlled friction overcomes the load controlled friction for strong adhesion parameter. We also observed that adhesion increases with a surface with a lower value of roughness. We need to extend this analysis to none IID distributions and also take into account the deformity of the asperities under pressure, changes in their geometry where the radius of curvature changes.

## 7-Acknowledgement

We are indebted to Daniel Bonn and Bart Weber for many detailed discussions on tribology.